\documentclass{aa} 

\usepackage{graphicx}
\usepackage{txfonts}
\usepackage{natbib}
\usepackage{url}
\bibpunct{(}{)}{;}{a}{}{,}

\def\apjl{ApJL}
\def\apjs{ApJS}

\newcommand{\Mpc}{$h^{-1}$\thinspace Mpc}

\newcommand{\vmh}{h^{-1}\mathrm{Mpc} }

\usepackage{color}

\begin{document}  

\title{Sloan Great Wall as a complex of superclusters with collapsing cores 
}

\author {Maret~Einasto\inst{1} 
\and Heidi Lietzen\inst{2}
\and Mirt~Gramann\inst{1} 
\and Elmo~Tempel\inst{1} 
\and Enn~Saar\inst{1,4}
\and Lauri~Juhan~Liivam\"agi\inst{1} 
\and Pekka~Hein\"am\"aki\inst{2}
\and Pasi~Nurmi\inst{2, 3}
\and Jaan~Einasto\inst{1,4,5} 
}

\institute{Tartu Observatory, Observatooriumi 1, 61602 T\~oravere, Estonia
\and
Tuorla Observatory, University of Turku, V\"ais\"al\"antie 20, Piikki\"o, Finland
\and
Finnish Centre for Astronomy with ESO (FINCA), University of Turku, 
V\"ais\"al\"antie 20, Piikki\"o, Finland
\and
Estonian Academy of Sciences, Kohtu 6, 10130 Tallinn, Estonia
\and
ICRANet, Piazza della Repubblica 10, 65122 Pescara, Italy
}

\authorrunning{Einasto, M. et al. }

\offprints{Einasto, M.}

\date{ Received   / Accepted   }

\titlerunning{SGW, collapsing}

\abstract
{
The formation and evolution of the cosmic web is governed 
by the gravitational attraction of  dark matter and antigravity  
of dark energy (cosmological constant). In the cosmic web, galaxy superclusters
or their high-density cores are the largest objects that may collapse
at present or during the future evolution. 
}
{
We  study the dynamical state and possible future evolution
of galaxy superclusters from the Sloan Great Wall (SGW), the richest galaxy system in the nearby
Universe. 
}
{
We calculated supercluster masses  using dynamical masses of galaxy groups 
and stellar masses of galaxies.
We employed normal mixture modelling to 
study the structure of rich SGW superclusters 
and search for components (cores) in superclusters.
We analysed the radial mass distribution in the high-density cores of superclusters 
centred approximately 
at rich clusters and 
used the spherical collapse model to study their dynamical
state. 
}
{
The lower limit of the total mass of the SGW is 
approximately $M = 2.5\times~10^{16}h^{-1}M_\odot$. 
Different mass estimators of superclusters agree well,
the main uncertainties in masses of superclusters 
come from missing groups and clusters. 
We detected three high-density cores in the richest SGW supercluster
(SCl~027) and two in the second richest supercluster  (SCl~019).
They have masses of $1.2 - 5.9 \times~10^{15}h^{-1}M_\odot$
and sizes of up to $\approx 60 $~\Mpc. 
The high-density cores of superclusters are very elongated, 
flattened perpendicularly to the line of sight. 
The comparison of the radial mass distribution 
in the high-density cores with the predictions of 
spherical collapse model suggests that
their central regions with radii smaller than
$8$~\Mpc\ and masses of up to $M = 2\times~10^{15}h^{-1}M_\odot$
may be collapsing. 
}
{ 
The rich SGW superclusters with their high-density  cores 
represent dynamically evolving environments 
for studies of the  properties of galaxies and galaxy systems.
}
%\end{abstract}


\maketitle

\section{Introduction} 
\label{sect:intro} 

Galaxy superclusters are the largest systems in the complex hierarchical network 
of galaxies, galaxy groups, clusters, and superclusters (the cosmic web).
The structure of the superclusters is formed during a hierarchical
evolution where the high-density cores of superclusters
are older and dynamically more evolved than outskirts regions. 
While full rich superclusters are not bound systems, 
their high-density cores
may collapse at present or in the course of the future evolution 
\citep{1998ApJ...492...45S, 2000AJ....120..523R, 2002AJ....124.1266R, 
2003NewA....8..439N, 2006A&A...447..133P, 2006MNRAS.366..803D, 
2011MNRAS.415..964L, 2014MNRAS.441.1601P, 2015A&A...575L..14C,
2015MNRAS.453..868O, 2015A&A...580A..69E, 2015A&A...581A.135G}.
This makes galaxy superclusters unique objects to study their properties
and the  properties and evolution of the galaxy systems
(groups, clusters, and filaments) inside the dynamically evolving environment
of superclusters.
The size, mass, and other properties of the high-density cores in superclusters
give us an information about the largest possibly collapsing objects in the Universe 
\citep{1991ApJ...374...29L, 2002MNRAS.337.1417G, 2003NewA....8..439N,
2015A&A...577A.144T, 2015ApJ...815...43L, 2015A&A...581A.135G}. 

The richest nearby galaxy system is the Sloan Great Wall (SGW), discovered in the Sloan Digital
Sky Survey 
\citep{2004ogci.conf....5V, 2005ApJ...624..463G}, which
consists of several rich and poor superclusters 
\citep{2011ApJ...736...51E}. 
\citet{2003A&A...405..821E, 2008ApJ...685...83E} noted that in the core region
of the richest supercluster in the SGW 
\citep[SCl~126 in their study, SCl~027 in]
[this notation is also used in our study]{2012A&A...539A..80L}
the concentration
of galaxy clusters in a sphere with diameter smaller than $10$~\Mpc\ is very
high. This region is a good candidate for a collapsing core of the supercluster.
The Sloan Great Wall is not fully covered by
the SDSS, its southern extension can be traced by the Las Campanas and 2dF
Redshift surveys \citep{2003A&A...405..821E,  2008ApJ...685...83E}. 
The SGW affects the measurements of the topology 
of the whole SDSS \citep{2005ApJ...633...11P, 2007MNRAS.374.1030S, 2008ApJ...675...16G}.
The analysis of  rich superclusters in the SGW has shown that
they have a different morphology
and galaxy and group content \citep{2007A&A...476..697E, 2010A&A...522A..92E, 
2011A&A...532A...5E, 2011ApJ...736...51E, 2014A&A...562A..87E}.
The richest supercluster in the SGW, SCl~027, is one of the
most elongated superclusters according to its overall shape \citep{2011A&A...532A...5E}. 
\citet{1998A&A...336...35J} have noted the flatness of this supercluster;
this supercluster is also aligned almost
perpendicular to the line of sight. The authors assumed that the flatness
of this supercluster is enhanced and we see the effect of
the matter inflow towards the supercluster axis,
in accordance to what we observe in the nearby space
for the Laniakea and Arrowheads superclusters 
\citep{2014Natur.513...71T, 2015ApJ...812...17P}.

The extreme observed objects like the SGW usually provide tests for theories.
For example, while \citet{2012ApJ...759L...7P} 
demonstrated that systems with sizes and richness similar to
the SGW can be reproduced in the $\Lambda$CDM model, \citet{2011MNRAS.417.2938S}
showed that systems as dense and massive as the SGW may be in tension with the
Gaussian initial conditions. Moreover, 
\citet{2007A&A...476..697E} found that the morphology of the richest supercluster
in the SGW is difficult to reproduce with simulations. 
\citet{2016A&A...587A.116E} showed that the superclusters from the SGW lie in a wall of
a shell-like structure around the rich cluster A1795 in the Bootes supercluster
with a radius of about $120 - 130$~\Mpc. 
The pattern of the cosmic web originates from processes
in the early Universe. However, it is not yet clear which processes 
cause shell-like structures in the local cosmic web.
This all motivates further studies of the properties of 
the SGW. 

The aim of the present paper is to determine the 
dynamical, total, and stellar masses of the SGW
superclusters, and to analyse the structure of rich SGW 
superclusters with normal mixture modelling.
We compare the mass distribution of the core regions of the superclusters
centred at rich galaxy clusters with the predictions of the 
spherical collapse model, which
describes the evolution of a 
spherically symmetric perturbation in an expanding universe. 
The dynamics of a collapsing shell is determined by the mass in its interior. 
The spherical collapse model has been discussed in detail by 
\citet{1980lssu.book.....P} 
\citep[see also references in ][]{2015A&A...581A.135G}.
We estimate
the dynamical state and possible future evolution
of the core regions of the SGW superclusters and 
discuss the possibility whether the high-density cores
of the SGW superclusters may merge into huge collapsing systems.

At \url{http://www.aai.ee/~maret/SGW.html}        
   we present an interactive 3D model that shows the distribution of
galaxy groups in the superclusters from the SGW.

We use  the following standard cosmological parameters below: 
the Hubble parameter $H_0=100~h$ km~s$^{-1}$ Mpc$^{-1}$,
the matter density $\Omega_{\rm m} = 0.27,$ and the 
dark energy density $\Omega_{\Lambda} = 0.73$ \citep{2011ApJS..192...18K}. 

\section{Data}
\label{sect:data} 

\begin{figure*}%[ht]
\centering
\resizebox{0.96\textwidth}{!}{\includegraphics[angle=0]{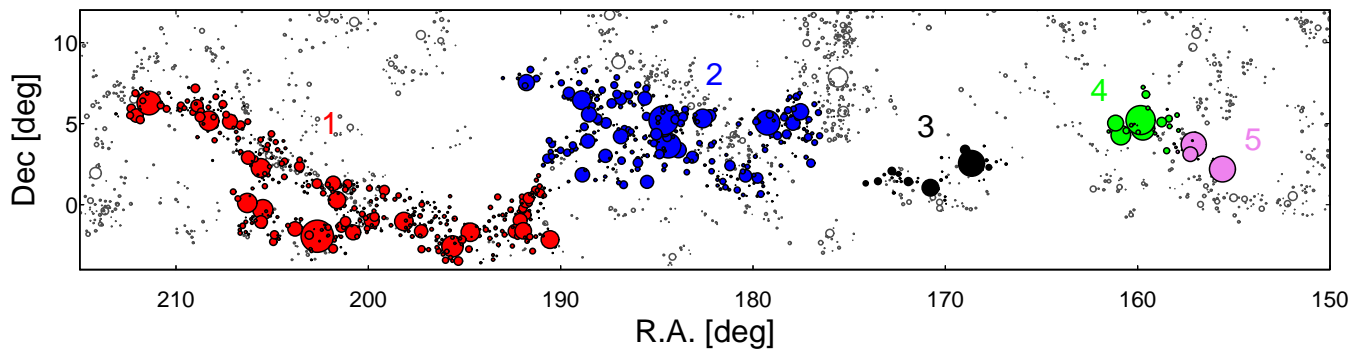}}
\caption{
Distribution of galaxy groups 
in the SGW superclusters in the sky plane in the redshift range 
$0.04 < z < 0.12$. Numbers are order numbers of superclusters
in Table~\ref{tab:scldata}, and different colours indicate
groups in these superclusters. Symbol sizes are proportional to the
size of groups in the sky plane. Grey symbols show galaxy groups in other 
superclusters in this redshift interval.
}
\label{fig:sclradec}
\end{figure*}

We selected data about superclusters and their galaxy and group content
from the supercluster and group catalogues by \citet{2012A&A...539A..80L} and  \citet{2012A&A...540A.106T, 2014A&A...566A...1T}. 
These catalogues are based on the MAIN sample of the eighth and tenth data release of the Sloan Digital Sky 
Survey \citep{2011ApJS..193...29A, 2014ApJS..211...17A} with 
apparent Galactic extinction-corrected $r$ magnitudes $r \leq 
17.77$ and redshifts $0.009 \leq z \leq 0.200$. 
We corrected the redshifts of galaxies with respect to our motion relative to the
CMB and computed the comoving distances  of galaxies  \citep[see][for details]{2002sgd..book.....M,
2014A&A...566A...1T}.
 
We used the luminosity density field to determine 
galaxy superclusters. 
In calculations of the
luminosity density field we applied the $1$~\Mpc\ step grid and 
the $B_3$ spline kernel at the smoothing length 8~\Mpc. 
While constructing the luminosity density field, we first suppressed the
redshift-space distortions (the so-called fingers of God) for groups as explained in 
\citet{2014A&A...566A...1T}.
Connected 
volumes above a certain density threshold were defined as superclusters. 
The calculation of the luminosity density field,
corrections for the faint galaxies missing from the catalogue,
 and determination of
superclusters are described in detail in   \citet{2012A&A...539A..80L}. 

The luminosity density field is a biased tracer of the underlying
mass field, as is shown, for example, by  the analysis of the mass-to-light ratios
of galaxy systems \citep{2014MNRAS.439.2505B, 2015A&A...580A..69E}. Therefore, 
in our analysis below we determine the masses of the high-density cores
of superclusters as described in Sect.~\ref{sect:masses} and 
do not directly use the luminosity density field for this purpose.

\citet{2011ApJ...736...51E} analysed the 
properties of the density field superclusters in the region of the
SGW at a series of density levels. 
They concluded that the density level $D8 = 5.0$
(in units of mean density, $\ell_{\mathrm{mean}}$ = 
1.65$\cdot10^{-2}$ $\frac{10^{10} h^{-2} L_\odot}{(\vmh)^3}$) is suitable
to determine individual superclusters.
At this density level superclusters in the SGW form separate systems,
and the SGW consists of two rich and three poor superclusters,  
at lower density levels they join into huge percolating systems
together with surrounding superclusters. In our study we use the data about
SGW superclusters chosen from the \citet{2012A&A...539A..80L} supercluster
catalogue at this density level.

If galaxies were moving solely with the general expansion of the Universe, 
the redshifts would accurately measure the radial distances of galaxies.
Since galaxies have peculiar velocities with respect to the general 
expansion, their redshift distances are distorted. 
The galaxy distribution in redshift space is different 
from that in real space on both small and large scales. 
On the largest scales, the amplitude of clustering is enhanced 
as a result of the coherent large-scale velocity field. 
In contrast to the elongation along the line of sight produced by incoherent 
velocities within galaxy groups, 
the galaxy distribution on large scales is flattened along the line of sight 
\citep[e.g. ][ and references therein]
{1987MNRAS.227....1K, 1994ApJ...425..382G, 1998ASSL..231..185H}.

In this paper we use the distribution of galaxies in redshift space. 
It would, of course, be preferable to define superclusters
as in \citet{2014Natur.513...71T} and  \citet{2015ApJ...812...17P}
considering the real distances of galaxies. 
Unfortunately, such data for the SGW region
are not yet available. Even for the local superclusters,
uncertainties are large \citep{2016AJ....152...50T}.
However, we can study the effect of peculiar velocities indirectly 
and compare the galaxy distribution in the sky plane 
and in the radial direction. 
For the richest SGW supercluster, SCl~027, 
the peculiar velocities may affect its overall shape, 
and we see the effect of the matter inflow towards the supercluster axis 
\citep{1998A&A...336...35J}.
In this paper we determine several components 
in rich SGW superclusters and study their sizes 
in the sky plane and in the radial direction in more detail.

\begin{table*}%[ht]
\caption{Data of individual superclusters in the SGW.}
\center
\begin{tabular}{rrrrrrrrrrrr} 
\hline\hline 
\multicolumn{1}{c}{(1)}&(2)&(3)&(4)& (5)&(6)&(7)&(8)&(9)&(10)&(11)&(12)\\      
\hline 
No. & Name &\multicolumn{1}{c}{ID}& $N_{\mathrm{gal}}$ & $N_{\mathrm{1}}$ 
& $N^{\mathrm{gr}}_{\mathrm{2-9}}$ & $N^{\mathrm{gr}}_{\mathrm{10}}$ & $\mathrm{Dist.}$  
 & $\mathrm{Diam.}$ & $D8_{\mathrm{max}}$ & $\mathrm{Vol.}$ &$L_{\mathrm{tot}}$  \\
\hline
1 & SCl~027 & 202-001+008 & 3222 & 706 & 381 & 50 & 255.6  & 107.0 & 14.0 & 25.9 & 51.6 \\
2 & SCl~019 & 184+003+007 & 2060 & 456 & 274 & 33 & 230.4  &  56.4 & 15.0 & 14.4 & 29.2 \\
3 & SCl~0499 & 168+002+007 & 408 &  60 & 26 & 7 & 227.7  &  34.1 &  7.5  & 2.0 & 4.77 \\
4 & SCl~0319 & 159+004+006 & 245 &  30 & 23 & 3 & 206.2  &  21.4 &  7.5  & 1.4 & 2.16 \\
5 & SCl~1109 & 157+003+007 & 120 &   4 &  5 & 3 & 219.2  &  12.1 &  5.2  & 0.2 & 1.49 \\
\hline
  & SGW &             &  6055 & 1256 & 709 & 96 &   &       &       & 43.9 & 89.22 \\
  
\hline
\label{tab:scldata}  
\end{tabular}\\
\tablefoot{
Columns are as follows:
(1) the order number of the supercluster; 
(2) the number of the supercluster in \citet{2012A&A...539A..80L};
(3) the supercluster ID AAA+BBB+ZZZ, where AAA is R.A., +/-BBB is Dec., and ZZZ is 100$z$;
(4) the number of galaxies in the supercluster, $N_{\mathrm{gal}}$;
(5) the number of single galaxies in the supercluster, $N_{\mathrm{1}}$;
(6) the number of groups with $2 - 9$ member galaxies, $N^{\mathrm{gr}}_{\mathrm{2-9}}$;
(7) the number of groups with $\geq 10$ member galaxies, $N^{\mathrm{gr}}_{\mathrm{10}}$;
(8) the distance of the supercluster (the
distance of the density maximum in the supercluster), in $h^{-1}$ Mpc;
(9) the supercluster diameter (the maximum distance between galaxies in
the supercluster), $\mathrm{Diam}$, in  $h^{-1}$ Mpc;
(10) the highest value of the luminosity-density field calculated with
the $8$~\Mpc\ smoothing kernel, $D8_{\mathrm{max}}$, in units of the mean luminosity density;
(11) the volume of the supercluster (the number of connected 3D grid cells in the 
luminosity density field, multiplied by the cell volume, $V$), 
in $10^{3}(h^{-1}$ Mpc$)^3$; 
(12) the total luminosity of the supercluster, calculated as the 
weighted total luminosity of galaxies in the supercluster, $L_{\mathrm{tot}}$, 
in  $10^{12}h^{-2} L_{\sun}$.
}
\end{table*}

Data about galaxy groups in  
superclusters were taken from the group catalogue 
by \citet{2014A&A...566A...1T}.
The  redshift-space distortions (fingers of God) for groups were supressed
as described in detail in \citet{2014A&A...566A...1T}.
Galaxy groups were  determined using 
the friends-of-friends cluster analysis 
method introduced in cosmology by \citet{1982Natur.300..407Z} and 
\citet{1982ApJ...257..423H}. A galaxy belongs to a 
group of galaxies if this galaxy has at least one group member galaxy closer 
than a linking length. In a flux-limited sample the density of galaxies slowly 
decreases with distance. To take this selection effect properly
into account
when constructing a group catalogue from a flux-limited sample, the 
linking length was rescaled with distance, calibrating the scaling relation by observed 
groups. As a result, the 
maximum sizes in the sky projection and the velocity dispersions of our groups 
are similar at all distances. The superclusters lie in a narrow distance interval, therefore we used groups from a flux-limited sample. The details about the data 
reduction, the group-finding procedure, and the description of the group catalogue can be found in 
\citet{2014A&A...566A...1T}.

In Table~\ref{tab:scldata} we list the data of the superclusters. 
The sky distribution of galaxy groups in the region covered by
superclusters from the SGW is shown in Fig.~\ref{fig:sclradec}.
The total length of the SGW is approximately $230$~\Mpc; without
poor superclusters, it is approximately $165$~\Mpc. This is comparable
with the estimate by \citet{2011MNRAS.417.2938S}, who found that the diameter
of the SGW is about $160$~\Mpc. With its length of $230$~\Mpc,\
the SGW is smaller than a recently discovered new member of the Wall family,
a very rich supercluster complex called the BOSS Great Wall
at a redshift of approximately $z = 0.47$
\citep[BGW, ][]{2016A&A...588L...4L}, which has a diameter of about $270$~\Mpc.
With its huge size and richness, the BGW is an even greater challenge to the
cosmological theories than the SGW.
In the catalogue of superclusters determined on the basis of X-ray
clusters by \citet{2013MNRAS.429.3272C}, the supercluster RXSCJ1305-0221
with its size of about $45$~\Mpc\ corresponds to the part of SCl~027 with
$Dec. < 2\degr$, centred approximately on the rich cluster A1650.  
The SGW is surrounded by voids. Closer to us, it is
located across the void behind the Hercules supercluster with
an approximate size of $120 - 140$~\Mpc. The distribution
of nearby rich superclusters was described in more detail in 
\citet{2011A&A...532A...5E} and in \citet{2016A&A...587A.116E}.

\section{Methods}
\label{sect:methods}

\subsection{Masses of superclusters}
\label{sect:masses} 

\begin{figure}%[ht]
\centering
\resizebox{0.445\textwidth}{!}{\includegraphics[angle=0]{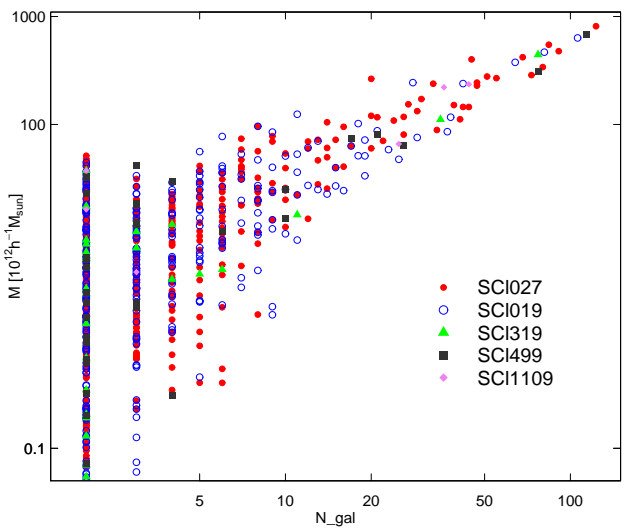}}
\caption{
Dynamical masses of groups vs. group richness for superclusters 
of the SGW.
}
\label{fig:nmass}
\end{figure}

To calculate the dynamical masses and mass-to-light ratios of superclusters, 
we used data about the dynamical masses  
of galaxy groups in superclusters from the catalogue of \citet{2014A&A...566A...1T}.
In Fig.~\ref{fig:nmass} we show the dynamical masses of all groups
in SGW superclusters. Masses are calculated  using the virial theorem, 
assuming that the galaxy velocity distribution 
and the Navarro-Frenk-White \citep[NFW, ][]{1997ApJ...490..493N}
projected density profile for galaxy distribution in the plane of the sky are symmetrical. 
For a detailed description of how the dynamical masses of groups were calculated
we refer to \citet{2014A&A...566A...1T}. The masses of poor groups are not well defined,
therefore we used the median values of group
masses instead of 
individual masses from \citet{2014A&A...566A...1T} for groups with 
$N_{\mathrm{gal}} = 2$ and $3$. 
To obtain a dynamical mass of a supercluster $M_{\mathrm{dyn}}$, 
we summed group dynamical masses. A similar procedure to calculate the dynamical
mass of the supercluster was used in \citet{2015A&A...580A..69E}.
The reliability of group mass estimation method 
and mass errors were analysed in 
\citet{2014MNRAS.441.1513O, 2015MNRAS.449.1897O}, 
where various mass estimation methods were tested on mock galaxy catalogues. 
The method used in this work performs reasonably well without 
a clear bias between true and estimated masses. 
Since in this paper we are only interested on the masses of superclusters 
(i.e. we sum over a large number of groups), 
the statistical uncertainty of an individual group mass estimate is not important. 
Errors in the supercluster dynamical masses that are due to the group mass errors  
are estimated to be about 0.3~dex. The supercluster total mass estimates
are dominated by systematic biases. Below we discuss some selection
effects that may affect mass estimates.

The total masses of superclusters
were calculated adding the estimated mass of faint groups and intercluster gas. 
Each region hosts some single galaxies. They may be the brightest galaxies
of faint groups in which other member galaxies are too faint to be 
observed within SDSS survey magnitude limits \citep{2009A&A...495...37T}.
We used the median mass of groups
with $2-3$ member galaxies in the SGW superclusters 
as the mass of these faint groups. 
To obtain the total mass of these faint groups, this median
mass was multiplied with the number of single galaxies. 
We also added 10\% of the total mass as the mass of  intercluster gas
\citep[see e.g. ][]{2016A&A...592A...6P}. 
Total masses of superclusters  $M^{\mathrm{g}}_{\mathrm{tot}}$ 
(here $g$ stands for both groups and gas)
were obtained by summing these mass estimates. 
Approximately 50\% of the supercluster total masses comes from groups with
at least ten member galaxies, 25\% of masses from poor groups
with fewer than ten galaxies, and 15\% of mass from faint groups
presented by single galaxies.

We also calculated total stellar masses of galaxies in superclusters
and stellar-to-total mass ratios using data about the stellar masses of 
supercluster member galaxies 
from the MPA-JHU spectroscopic catalogue \citep{2004ApJ...613..898T, 
2004MNRAS.351.1151B} in the SDSS CAS database. 
In this catalogue the different properties of 
galaxies are obtained by fitting SDSS photometry and spectra with
the stellar population synthesis models developed by \citet{2003MNRAS.344.1000B}.
The stellar masses of galaxies are estimated from the 
galaxy photometry \citep{2003MNRAS.341...33K}. 
The stellar mass of superclusters $M^{\mathrm{*}}$ is the sum of the stellar
masses of galaxies in a supercluster. The stellar-to-total
mass ratio $M^{\mathrm{*}}$/$M^{\mathrm{g}}_{\mathrm{tot}}$ is the ratio of the
stellar mass of galaxies, $M^{\mathrm{*}}$, and total mass $M^{\mathrm{g}}_{\mathrm{tot}}$
of a supercluster.

In addition, data about stellar masses of the most luminous  galaxies in a group 
can be used to calculate the 
mass of the haloes to which galaxies belong, employing
the relation of stellar mass $M_{\mbox{*}}$ to halo mass $M_{\mbox{halo}}$ from \citet{2010ApJ...710..903M},
\begin{equation}
\frac{M_{*}}{M_{\mbox{halo}}}=2\left(\frac{M_{*}}{M_{\mbox{halo}}}\right)_0 
\left[\left(\frac{M_{\mbox{halo}}}{M_1}\right)^
{-\beta}+\left(\frac{M_{\mbox{halo}}}{M_1}\right)^\gamma\right]^{-1},
\label{eq:stmass}
\end{equation}
where $(M_{\mbox{*}}/M_{\mbox{halo}})_0=0.02817$ 
is the normalisation of the stellar to halo mass relation, 
the halo mass $M_{\mbox{halo}}$ is the virial mass of haloes, 
$M_1 = 7.925\times~10^{11}M_\odot$
is a characteristic mass,
and $\beta=1.068$ and $\gamma=0.611$ 
are the slopes of the low- and high-mass ends of the relation, respectively. 
This method was used by \citet{2016A&A...588L...4L} to estimate the mass
of the BGW superclusters. 
We used this mass estimate for single galaxies and groups of
up to nine
member galaxies and dynamical masses of groups from the group
catalogue of \citet{2014A&A...566A...1T}
for richer groups to calculate a  supercluster mass
estimate from stellar masses of galaxies $M^{\mathrm{*}}_{\mathrm{tot}}$.
To make this mass estimate comparable to the total mass of superclusters
calculated using group dynamical masses, $M^{\mathrm{g}}_{\mathrm{tot}}$,
we also added 10\% of the total mass as the mass of gas and denote this mass as $M^{\mathrm{*g}}_{\mathrm{tot}}$.

\subsection{Multidimensional normal mixture modelling}
\label{sect:mclust} 

We studied the structure of rich superclusters from the SGW with
multidimensional normal mixture modelling, which  is based
on the analysis of a finite mixture of multivariate Gaussian distributions. 
Each component in the distribution corresponds to a separate component in the
data.  For this analysis we employed
the {\it Mclust} package \citep{fr2002, fr2012}
in statistical R environment \citep{ig96} for the model-based clustering analysis.
{\it Mclust} searches for an optimal model for the clustering of the data
among the models with varying shape, orientation, and volume,
and finds the optimal number of  components in the data and the membership
of components (classification of the data). 
{\it Mclust} also calculates
the uncertainty of the classification and determines for each datapoint 
the probability of belonging to a component. 
The uncertainty of classification is defined as one minus the 
highest probability of a datapoint to belong to a component. 
The mean uncertainty
for the full sample  is a statistical estimate of the reliability
of the results.
To  study the structure of superclusters, 
the input data for {\it Mclust} were 
the coordinates of galaxy groups in superclusters.
This package has been used, for example, to search for substructure
in galaxy groups and clusters and to refine the group-finding
algorithm \citep{2010A&A...522A..92E, 2012A&A...540A.123E, 
2013MNRAS.434..784R, 2016A&A...588A..14T}, to analyse the distribution
of different galaxy populations in superclusters \citep{2011ApJ...736...51E},
and in morphological classification of superclusters \citep{2011A&A...532A...5E}.

\begin{table*}%[ht]
\setlength{\tabcolsep}{4pt}
\caption{Supercluster masses.}
\center
\begin{tabular}{rrrrrrrrr} 
\hline\hline 
(1)&(2)&(3)&(4)& (5)&(6)&(7)&(8)&(9)\\      
\hline 
No. & Name & $M_{\mathrm{dyn}}$ & $M^{\mathrm{g}}_{\mathrm{tot}}$ & $M^{\mathrm{*}}$ 
&  $M^{\mathrm{*}}$/$M^{\mathrm{g}}_{\mathrm{tot}}$ & $M^{\mathrm{*}}_{\mathrm{tot}}$&
$M^{\mathrm{*g}}_{\mathrm{tot}}$  
& $M^{\mathrm{g}}_{\mathrm{tot}}/L$  \\
& & $10^{15}h^{-1}M_\odot$ & $10^{15}h^{-1}M_\odot$ & $10^{13}h^{-1}M_\odot$ & &
$10^{15}h^{-1}M_\odot$ & $10^{15}h^{-1}M_\odot$ & $h\,M_\odot$/$L_\odot$ \\
\hline
1 & SCl~027 &   10.41 & 14.00 &  16.61 & 0.012 & 11.13 & 12.24   & 271  \\
2 & SCl~019 &    5.42 &  7.03 &   9.09 & 0.013 &  6.41 &  7.05   & 241  \\
3 & SCl~0499 &   1.41  & 1.74 &   1.69 &  0.010&  1.15 &  1.27   & 365  \\
4 & SCl~0319 &   0.66  & 0.82 &   0.87 &  0.011&  0.76 &  0.84   & 380  \\
5 & SCl~1109 &   0.56  & 0.63 &   0.69 &  0.011&  0.29 &  0.32   & 423  \\
& SGW    &   18.46 & 24.22&  28.95 &  0.012& 19.74 & 21.71   & 272  \\ 
\hline
\label{tab:masses}  
\end{tabular}\\
\tablefoot{
Columns are as follows:
(1) the order number of the supercluster;
(2) the ID of the supercluster in \citet{2012A&A...539A..80L};
(3) the dynamical mass of the supercluster, $M_{\mathrm{dyn}}$; %, in $10^{15}h^{-1}M_\odot$; 
(4) the total mass of the supercluster (including faint groups and intercluster
    gas, see text), $M^{\mathrm{g}}_{\mathrm{tot}}$; 
(5) the stellar mass of the supercluster, $M^{\mathrm{*}}$; 
(6) $M^{\mathrm{*}}$/$M^{\mathrm{g}}_{\mathrm{tot}}$; 
(7) the total mass of the supercluster, $M^{\mathrm{*}}_{\mathrm{tot}}$, 
calculated using the stellar mass - halo mass relation (see text); %, in $10^{15}h^{-1}M_\odot$; 
(8) the total mass of the supercluster, $M^{\mathrm{*g}}_{\mathrm{tot}}$, 
calculated using adding gas mass estimate to the mass $M^{\mathrm{*}}_{\mathrm{tot}}$;
%, in $10^{15}h^{-1}M_\odot$; 
(9) the mass-to-light ratio, $M^{\mathrm{g}}_{\mathrm{tot}}/L$. %, in $h\,M_\odot$/$L_\odot$. 
}
\end{table*}

\subsection{Spherical collapse model}
\label{sect:coll} 

The spherical collapse model describes the evolution of a spherical 
perturbation in an expanding universe. 
This model was studied by
\citet{1980lssu.book.....P},
\citet{1984ApJ...284..439P}, \citet{1991MNRAS.251..128L},
\citet{1996MNRAS.282..263E}, and \citet{2001idm..conf..121L}.  
In the standard models with cosmological constant, 
the dark energy started accelerating the expansion at the redshift 
$z \approx 0.7$ 
and the formation of structure slowed down. 
At the present epoch, the largest bound structures are just forming.  
In the future evolution of the universe, 
these bound systems separate from each other at an accelerating rate, 
forming isolated "island universes"
\citep{2002PhRvD..65l3518C, 2003ApJ...596..713B, 2006MNRAS.366..803D}.

\citet{2015A&A...575L..14C} and \citet{2015A&A...581A.135G}  analysed 
the characteristic density contrasts for the turnaround 
and future collapse in different spherical collapse models. 
These density contrasts can be used to derive the relations 
between radius of a perturbation and the interior mass for each essential epoch.

For a spherical volume $V = 4\pi R^3/3 $ 
with radius $R$ 
the density ratio to the mean density (overdensity)
$\Delta\rho = \rho/\rho_{\mathrm{m}}$ can be calculated as
\begin{equation}
\Delta\rho=6.88\,\Omega_\mathrm{m}^{-1}\left(\frac{M}{10^{15}h^{-1}M_\odot}\right)
        \left(\frac{R}{5h^{-1}\mathrm{Mpc}}\right)^{-3}.
\label{eq:sph}
\end{equation}
From Eq.~(\ref{eq:sph}) we can find the mass of a structure as
\begin{equation}
M(R)=1.45\cdot10^{14}\,\Omega_\mathrm{m}\Delta\rho\left(R/5h^{-1}\mathrm{Mpc}\right)^3h^{-1}M_\odot.
\label{eq:mass1}
\end{equation}

%%%%%%%%%%%%

{\it Turnaround.} One essential moment in the evolution of a spherical perturbation 
is called turnaround, the moment when the sphere stops expanding 
together with the universe and the collapse begins. 
At the turnaround, the perturbation decouples 
entirely from the Hubble flow of the homogeneous background. 
The spherically averaged radial velocity around a system 
in the shell of radius $R$ can be written as $u = HR - v_{\mathrm{pec}}$, 
where $v_{\mathrm{H}} = HR$ is the Hubble expansion velocity and 
$v_{\mathrm{pec}}$ is 
the averaged radial peculiar velocity towards the centre of the system. 
At the turnaround point, the peculiar velocity $v_{\mathrm{pec}} = HR$ and $u = 0$. 
The peculiar velocity $v_{\mathrm{pec}}$ 
is directly related to the overdensity $\Delta\rho$.
For $\Omega_{\mathrm{m}} = 0.27$ and $\Omega_{\mathrm{\Lambda}} = 0.73$  
the overdensity at the turnaround point in the spherical collapse model
is  $\Delta\rho_{\mathrm{T}} =  13.1$ and  
the mass of a structure  at the turnaround point is \citep{2015A&A...581A.135G}
\begin{equation}
M_\mathrm{T}(R)=5.1\cdot10^{14}\left(R/5h^{-1}\mathrm{Mpc}\right)^3h^{-1}M_\odot.
\label{eq:mrt}
\end{equation}

%%%%%%%%%%%%%%%%

{\it Future collapse.} The superclusters that have not reached the turnaround at present may 
eventually turnaround and collapse in the future \citep{2006MNRAS.366..803D}. 
\citet{2015A&A...575L..14C} showed that for $\Omega_{\mathrm{m}} = 0.27$
the overdensity  
for the future collapse 
 $\Delta\rho_{\mathrm{FC}} =  8.73,$ which gives
the minimum mass of the structure that will 
turn around and collapse in the future as
\begin{equation}
M_\mathrm{FC}(R)=3.4\cdot10^{14}\left(R/5h^{-1}\mathrm{Mpc}\right)^3h^{-1}M_\odot.
\label{eq:mrfvs}
\end{equation}

%%%%%%%%%

The spherical collapse model 
has been applied to study, for example, 
high-density cores in the Corona Borealis supercluster 
\citep{1998ApJ...492...45S, 2014MNRAS.441.1601P}, 
in the Shapley supercluster \citep{2000AJ....120..523R, 2006A&A...447..133P, 2015A&A...575L..14C},
in the A2199 supercluster \citep{2002AJ....124.1266R}, and in the A2142 supercluster
\citep{2015A&A...580A..69E, 2015A&A...581A.135G}.

Below we analyse the structure of superclusters, find their 
components, and  study their masses, sizes, and shapes. 
We analyse the observed radial mass distribution in components
centred on rich galaxy clusters and compare it with
the turnaround and future collapse masses predicted by the spherical
collapse model.

\section{Results}
\label{sect:results} 

\subsection{Masses and mass-to-light ratios of the SGW superclusters}
\label{sect:mscl}

All mass estimates for superclusters are given in Table~\ref{tab:masses},
which shows that 
for the rich SGW superclusters
different mass estimators give quite  close 
values of mass.
For the supercluster SCl~019 they are almost identical. The differences between
masses from dynamical masses of groups and stellar masses of galaxies are the largest
for two poor superclusters, SCl~499 and SCl~1109,
where the number of groups is rather low (especially in SCl~1109) and mass estimates
have larger scatter.
The difference between $M^{\mathrm{g}}_{\mathrm{tot}}$ and $M^{\mathrm{*g}}_{\mathrm{tot}}$
comes from the difference of how the group masses are estimated for poor  
(with fewer than ten member galaxies) groups.

\begin{table*}%[ht]
\setlength{\tabcolsep}{4pt}
\caption{Data of the SCl~027 components.}
\center
\begin{tabular}{rrrrrrrrrrrrr} 
\hline\hline  
(1)&(2)&(3)&(4)& (5)&(6)&(7)&(8)&(9)&(10)&(11)&(12)&(13)\\      
\hline 
 $Nr$ & $N_{\mathrm{gr}}$ & $N_{\mathrm{10}}$ & $N_{\mathrm{1}}$ 
 & $M_{\mathrm{dyn}}$ &  $M^{\mathrm{g}}_{\mathrm{tot}}$ & $dx$ & $dy$ & $dz$ & $dx/dy$ & $dz/dy$ & $D8$& ID\\
 & && & [$10^{15}h^{-1}M_\odot$]& [$10^{15}h^{-1}M_\odot$] &$h^{-1}$ Mpc&$h^{-1}$ Mpc&$h^{-1}$ Mpc & & & \\ 
\hline
1 & 218 &  9 & 132 & 2.4 & 2.7 & 14.4& 40.0 & 29.7& 0.4& 0.7& 7.2& \object{A1773}  \\
2 &  71 &  5 &  41 & 1.0 & 1.1 & 24.9& 33.5 & 10.7& 0.7& 0.3& 5.7&   \\
3 & 451 & 18 & 270 & 5.4 & 5.9 & 20.3& 57.3 & 21.9& 0.4& 0.4& 7.2& \object{A1650}  \\
4 & 115 &  8 &  60 & 1.3 & 1.4 & 29.1& 12.2 & 29.5& 2.4& 2.4& 6.0&   \\
5 & 184 & 10 & 120 & 2.2 & 2.4 & 18.5& 30.2 & 18.8& 0.6& 0.6& 7.5& \object{A1750}  \\
\hline                
\label{tab:comp27}  
\end{tabular}\\
\tablefoot{
Columns are as follows:
(1): Component number;
(2): the total number of groups in the component;
(3): the number of groups with at least ten galaxies in the component;
(4): the number of single galaxies in the component;
(5): the dynamical mass of the component;
(6): the total mass of the component (including intercluster gas);
(7--9): largest extent of the component along the $dx$, $dy$, and $dz$ direction;
(10--11) the ratio of $x/y$ and $z/y$ axes;
(12) median value of the luminosity density in the component;
(13) ID of the central cluster in the component.
}
\end{table*}

%%%%%%%%%%%%%%%%%%%5

The main sources of mass uncertainties are 
several selection effects that may affect the supercluster mass estimates.
For example, the SDSS galaxy sample is incomplete because of fibre 
collisions: the smallest separation between spectroscopic fibres is 55",
and about 6\% of galaxies in the SDSS are without observed spectra. 
\citet{2012A&A...540A.106T} studied the effect of missing galaxies
on a group catalogue and concluded that this mostly affects galaxy pairs.
The authors estimated that approximately 8\% of galaxy pairs may be missing from the catalogue.
Since they are included as single galaxies and we take them into account as
the main galaxies of faint groups, the effect of fibre collisions to
our results is minor.

Some galaxy groups and clusters are missing from the supercluster SCl~027
since its southern extension is not covered by SDSS survey. 
One missing rich cluster is the cluster A1651 \citep{2003A&A...405..821E, 2008ApJ...685...83E}.
We may assume that the  mass of this cluster
is of the order of the  mass of A1650 with dynamical
mass of 
$M_{\mathrm{dyn}} \approx 0.3\times~10^{15}h^{-1}M_\odot$.
Then the dynamical mass of SCl~027 becomes 
$M_{\mathrm{dyn}} = 10.7\times~10^{15}h^{-1}M_\odot$,
the total mass of SCl~027 is then 
$M^{\mathrm{g}}_{\mathrm{tot}} = 14.3\times~10^{15}h^{-1}M_\odot$,
and the total mass of the SGW is 
$M^{\mathrm{g}}_{\mathrm{tot}} = 24.5\times~10^{15}h^{-1}M_\odot$.
These masses are lower limits only, since poor groups and single
galaxies are also missing from the supercluster.

We used data of the superclusters from the supercluster catalogue with a fixed
luminosity density limit, $D8 = 5$, and may miss lower density outskirts of the
superclusters. Our analysis, focused on the high-density cores
regions of superclusters, is not affected by this choice, but it may
lead to an underestimation of the total masses of superclusters.

Mass-to-light ratios $M/L$ in Table~\ref{tab:masses}
for the rich superclusters and for the full
SGW are $M/L < 300$~$h\,M_\odot$/$L_\odot$, approximately the same as
the mass-to-light ratio
of the supercluster A2142 with $M/L = 287$~$h\,M_\odot$/$L_\odot$ \citep{2015A&A...580A..69E}.
The values of the mass-to-light ratios 
for poor superclusters in the SGW depend
on the mass estimators that have larger scatter (Table~\ref{tab:masses})
than those for rich superclusters.

\subsection{Structure and mass distribution in rich SGW superclusters}
\label{sect:str}

To study the structure of the rich SGW superclusters,
we searched for
possible components (cores) in superclusters with normal mixture modelling. 
As an input for calculations we
used the Cartesian coordinates of galaxy groups in superclusters
(including single galaxies), 
defined as $x = d\cos\delta\cos\alpha_{\mathrm{scl}}$,
$y = d\cos\delta\sin\alpha_{\mathrm{scl}}$, and $z = d\sin\delta$, 
where $d$ is the comoving distance, 
$\alpha_{\mathrm{scl}} = \alpha - \alpha_{\mathrm{mean}}$ 
is the right ascension (centred on the supercluster mean, 
$\alpha_{\mathrm{mean}}$), and $\delta$ is 
the declination of the group centre. 
The angle between the $x$ coordinate and the line of sight
is smaller than $5$ degrees, so that we can
consider $x$ direction as the line-of-sight direction.
We analysed the masses, sizes, and shapes of the components.
Next we study the group content and radial mass distribution 
in the cores of rich SGW superclusters centred on rich galaxy clusters
and compare it
with the predictions of the  spherical collapse model. 

\subsubsection{Supercluster SCl~027}
\label{sect:scl027}

\begin{figure}%[ht]
\centering
\resizebox{0.45\textwidth}{!}{\includegraphics[angle=0]{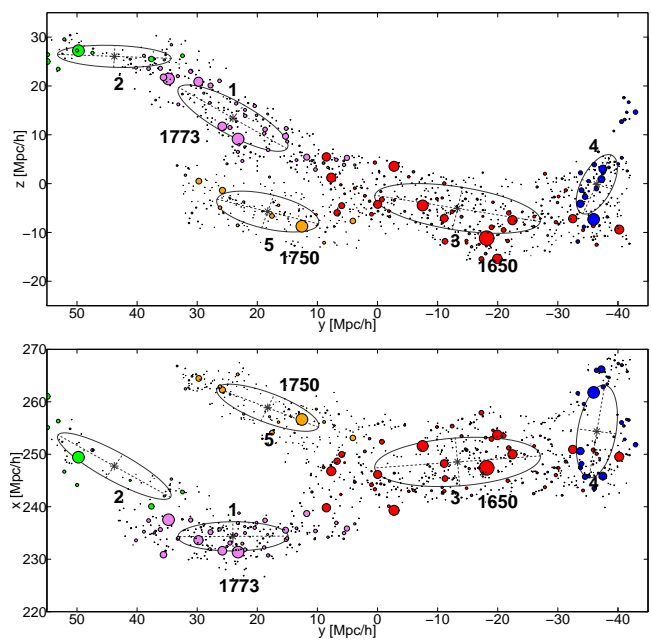}}
\caption{
Distribution of galaxy groups in SCl~027 in Cartesian coordinates.   
Upper panel: $yz,$ and lower panel: $yx$ plane. 
Filled circles of different colours correspond to galaxy groups from different 
components. The size of the symbols is proportional to the size of the groups
in the sky plane. Ellipses are $1\sigma$ covariance ellipses of components as found by 
normal mixture modelling. Numbers show order numbers of components from 
Table~\ref{tab:comp27} and Abell numbers of rich clusters in the high-density
cores of superclusters.
}
\label{fig:mclust27}
\end{figure}

Normal mixture modelling  with 
{\it Mclust} revealed five components in the supercluster SCl~027. 
The components were detected with 
uncertainties of $0.003$, which shows that
the results of the modelling has very high significance. 
In Fig.~\ref{fig:mclust27} we show the distribution
of galaxy groups (including faint groups represented by single galaxies)
in different components of SCl~027. In Table~\ref{tab:comp27} we 
list  data about components
(the number of groups, masses, sizes, and median luminosity density).

The supercluster SCl~027 has
three rich components, approximately centred
on rich galaxy clusters (A1650, A1750, and A1773, see Table~\ref{tab:comp27}). 
Each high-density core also hosts other rich galaxy clusters.
These compnents have total masses of  $2.4 - 5.9\times~10^{15}h^{-1}M_\odot$
and a largest extent of  $30-57$~\Mpc\ (Table~\ref{tab:comp27}).
Their median luminosity density is $D8 > 7$.
The same luminosity density limit was found for the high-density cores of
superclusters in \citet{2007A&A...464..815E}.
These components are the high-density cores of the supercluster. 

SCl~027  also has two less massive components with total masses
lower than $1.5\times~10^{15}h^{-1}M_\odot$. 
Their median luminosity density is $D8 \leq 6$.
These components form outlying branches  of the supercluster. 
Figure~\ref{fig:mclust27} shows that the poor component 4 is 
probably a combination of two very poor components that
coincide in the sky plane (upper panel), 
but are separated in velocity space (lower panel).

Figure~\ref{fig:mclust27} and Table~\ref{tab:comp27} show that
the high-density components in SCl~027  are very elongated along the  $y$ axis
and have a largest extent of  $30-57$~\Mpc. 
The extent of the components along the  $x$-axis direction (along the line of sight)
is $14-20$~\Mpc. The rich components 1 and 3 lie across the line
of sight.

\begin{table}%[ht]
\setlength{\tabcolsep}{4pt}
\caption{Masses and radii of the central regions of the SCl~027 components.}
\center
\begin{tabular}{rrrrrr} 
\hline\hline  
(1)&(2)&(3)&(4)&(5)&(6)\\      
\hline 
 $Epoch$ & $N_{\mathrm{gr}}$ & $N_{\mathrm{1}}$ & $M_{\mathrm{dyn}}$ &  $M^{\mathrm{g}}_{\mathrm{tot}}$ & $R$ \\
  && & [$10^{15}h^{-1}M_\odot$]& [$10^{15}h^{-1}M_\odot$] &$h^{-1}$ Mpc\\ 
\hline
\multicolumn{2}{l}{A1650 region} &&&&\\
 $T$   & 11 & 13   & 0.62& 0.68  & 5.5 \\
 $FC$  & 32 & 55   & 1.31& 1.44  & 8.0 \\
\hline                
\multicolumn{2}{l}{A1750 region} &&&&\\
 $T$   & 19 & 29   & 1.16& 1.28  & 6.5 \\
 $FC$  & 27 & 52   & 1.27& 1.40  & 8.0 \\
\hline                
\multicolumn{2}{l}{A1773 region} &&&&\\
 $T$   & 17 & 28   & 0.89& 0.98  & 6.0 \\
 $FC$  & 26 & 54   & 0.98& 1.08  & 7.5 \\
\hline
\label{tab:D8prop27}  
\end{tabular}\\
\tablefoot{
Columns are as follows:
(1): Epoch (T: turnaround, FC: future collapse);
(2): the number of groups in a region;
(3): the number of single galaxies in a region;
(4): the dynamical mass of a region;
(5): the total mass of a region (including intercluster gas, see text);
(6): the radius of the region.
}
\end{table}

\begin{figure}%[ht]
\centering
\resizebox{0.445\textwidth}{!}{\includegraphics[angle=0]{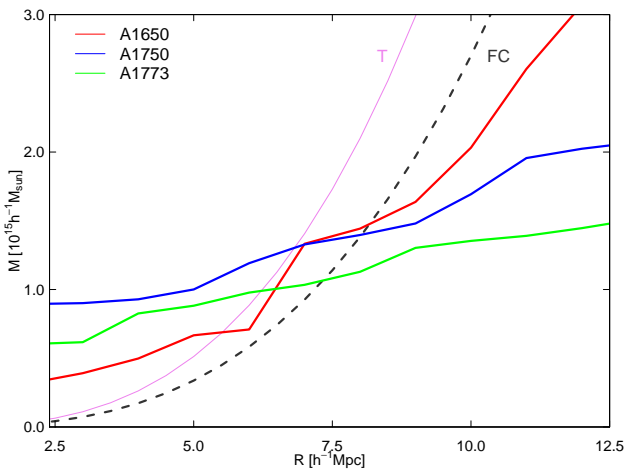}}
\caption{
Mass-radius relation for the high-density cores 
in the supercluster SCl~027. 
Violet line show embedded mass $M$ versus radius of a sphere $R$ 
for the turnaround $T$, and dashed grey line for the future collapse  $FC$.
The red line shows embedded mass versus radius for the region around the cluster A1650, 
the blue line shows the region around the cluster A1750, and the green
line the region around the cluster A1773.
}
\label{fig:mr27}
\end{figure}

Next we analyse the 
mass distribution in the high-density core regions (components 1, 3, and 5)
of SCl~027. 
We found galaxy groups and single galaxies in the spheres
of an increasing radius around rich clusters 
and calculated the embedded mass.
In Fig.~\ref{fig:mr27} we plot the embedded mass $M(R)$ 
versus radius of a sphere $R$ for these cores of SCl~027 and  
show the theoretical $M-R$ curves
from the spherical collapse model for comparison. According to this model,
the radii where the observed mass-radius lines cross the turnaround
and future collapse lines correspond to  maximum sizes of the regions that
may collapse now or in the future. 
Objects in the area above the turnaround line 
may have reached the turnaround and are already collapsing, and
objects above the future collapse line may collapse in the future. 
We present in 
Table~\ref{tab:D8prop27} 
the number of galaxy groups and single galaxies in spheres whose
radius corresponds to these essential epochs of the collapse
together with the embedded masses.

{\it A1650 region.} 
The central cluster in this region,  
\object{A1650}, is an X-ray
cluster \citep[see ][for details]{2011A&A...532A...5E}. 
The rich cluster \object{A1663} also lies in the central region of the component
with a distance of $6$~\Mpc\ from A1650.
One rich cluster close to this region, 
the X-ray cluster \object{A1651,} lies beyond
the SDSS survey declination limits
\citep{2003A&A...405..821E, 2008ApJ...685...83E}. 
The mass of the region within the turnaround radius of A1650 may be underestimated
because of the mass of this cluster. Assuming that A1651 
is a member of this high-density region, the total mass of the region increases
approximately by $0.3\times~10^{15}h^{-1}M_\odot$, and the radius 
may become approximately $10$~\Mpc.
In this case, another Abell cluster, \object{A1620,} with a mass of $0.4\times~10^{15}h^{-1}M_\odot$
and at a distance from A1650 of approximately
 $10$~\Mpc,\ may join the future collapse region 
around A1650. The total mass of the region within the future 
collapse radius may then be at least 
$2.1\times~10^{15}h^{-1}M_\odot$, which is approximately
one-third of the total mass of this high-density core. 

{\it A1750 region.} 
The component A1750 forms a separate branch in the supercluster
with a length of about $30$~\Mpc\ (Table~\ref{tab:comp27}).
The richest galaxy cluster here is \object{A1750} with the highest mass of the
SGW clusters, approximately $0.8\times~10^{15}h^{-1}M_\odot$. 
This is a multicomponent X-ray cluster that shows
signs of past merging \citep[][and references therein]{2004A&A...415..821B, 2010A&A...522A..92E}. 
The comparison of the radial mass distribution around the cluster
A1750 with the predictions of the spherical collapse model
shows that the central part of the component
with a maximum radius of $8$~\Mpc\ is collapsing or will collapse
in the future. 
Merging events in A1750 may be enhanced by the collapse of the whole region.

{\it A1773   region.} 
The Abell cluster \object{A1773}  with a mass of 
$0.5\times~10^{15}h^{-1}M_\odot$ 
is the most massive cluster in this high-density core.
The mass distribution around A1773 shows that the region around it
within a radius of about $6$~\Mpc\ may be collapsing.
Another rich cluster (\object{A1809} with a mass of $0.3\times~10^{15}h^{-1}M_\odot$)
lies at a distance of about $17$~\Mpc\ from A1773 
at the edge of the component, beyond the collapsing region. 

To summarise, according to the comparison
of the mass distribution with the predictions of the spherical collapse model,
the central parts of these 
components in SCl~027 with sizes smaller than $6.5$~\Mpc\ 
have already reached the turnaround and started
to collapse. The sizes of possible future collapse regions do not exceed
$8$~\Mpc. 
In this model, collapsing regions are surrounded by regions in the supercluster
that will not collapse and continue to expand.   
The cores in the 
supercluster are very elongated, with a largest extent of
$30-57$~\Mpc. We may assume that the spherical collapse model only describes
the central parts of the possibly collapsing regions,
and the actual size of collapsing regions 
across the line of sight may be larger. 
The flattened shape of the supercluster cores  
may be a signature of a possible collapse.
However, within
collapsing regions and also 
beyond the turnaround regions the group redshifts 
are poor indicators of their distance
\citep[see also the discussion in ][]{2007MNRAS.376.1577D}.
This affects the estimation of the sizes and masses of the possibly
collapsing regions. 
\citet{2006MNRAS.366..803D} showed that the
spherical collapse model applied in real space
overestimates the mass of the collapsing structure, and this may lead 
to the overestimation of the sizes of collapsing regions. 
The redshift corrections for regions beyond the turnaround region 
may lead to the underestimation of the sizes of the regions
\citep{2007MNRAS.376.1577D}. 
Without knowing the real distances of galaxy groups in the regions
it is very difficult to give precise values of the sizes
and other parameters of the collapsing regions.
This shows the limitations of the spherical collapse model
and also the need for correct distances of galaxy groups.

Approximately $60-75$~\% of the total mass in (now or in the future)
collapsing regions in SCl~027 comes from the mass of  galaxy groups with
at least ten member galaxies. This is slightly higher than in the 
full supercluster ($50$~\%, see Sect.~\ref{sect:masses}).
All regions contain  single galaxies that represent faint galaxy groups.
We found that approximately $6-9$\% of the total mass in the turnaround
regions and $12-14$\% of the total mass in the future collapse regions 
comes from these faint groups, showing that the fraction of
single galaxies decreases towards the region centres. This may be due to the
selection effect: it is possible that the group-finding algorithm
in \citet{2014A&A...566A...1T} adds single galaxies near rich groups
to the clusters. In this case, the masses of the faint groups have been taken into 
account when calculating the masses of rich groups. 
During the future evolution, these groups and galaxies may merge with the
main cluster of a region, as is shown in the analysis of the
future evolution
of superclusters from simulations \citep{2009MNRAS.399...97A}.
It is therefore also possible that the lower fraction of single galaxies
in the turnaround regions is evidence of the merging of single
galaxies and poor groups with the main cluster during collapse.
When we overestimated the mass of faint groups,
these fractions show that  errors of masses of the regions due to this are 
smaller than $10$\%.
The mass-to-light ratios $M/L$ of the collapsing regions are 
$280-300$$h\,M_\odot$/$L_\odot$, 
the same as the $M/L$  for the full supercluster. 

The ratio of stellar masses to the total mass in the collapsing
regions is lower than in the supercluster on average, $\approx 0.008-0.009$.
This ratio increases towards lower halo masses \citep{2010MNRAS.407..263A,
2014MNRAS.439.2505B, 2015ApJ...799L..17P}. This may be the reason
why this ratio has a lower value in the high-density cores than
in the full supercluster.

\subsubsection{Supercluster SCl~019}
\label{sect:scl019} 

\begin{figure}%[ht]
\centering
\resizebox{0.45\textwidth}{!}{\includegraphics[angle=0]{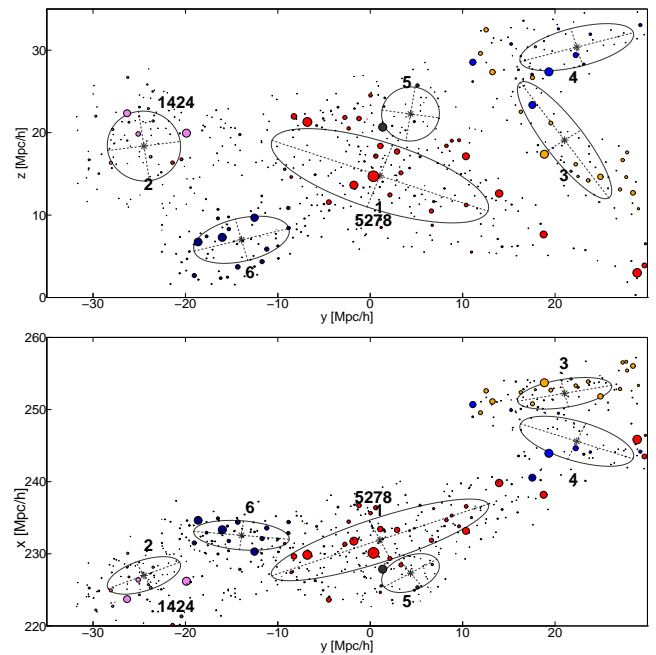}}
\caption{
Distribution of galaxy groups in SCl~019 in Cartesian coordinates.   
Upper panel: $yz,$ and lower panel: $yx$ plane. 
Filled circles of different colours correspond to galaxy groups from different 
components. The size of the symbols is proportional to the size of groups
in the sky plane. Ellipses are $1\sigma$ covariance ellipses of components as found by 
normal mixture modelling. Numbers show order numbers of components from 
Table~\ref{tab:comp19} and numbers of rich clusters in the high-density
cores of the supercluster.
}
\label{fig:mclust19}
\end{figure}

In SCl~019 the analysis with {\it Mclust}
identified six components. The uncertainty of the classification is $0.004$,
showing that the components in the supercluster have been found with very
high significance. The distribution
of galaxy groups (including faint groups represented by single galaxies)
in the components is shown in Fig.~\ref{fig:mclust19}.
Table~\ref{tab:comp19} presents
the number of groups, masses, and sizes of supercluster components.  

Two components in SCl~019 have a median luminosity density of 
$D8 \geq 7$, they are the high-density cores of the supercluster.
The total mass in these components is $1.2 - 3.2\times~10^{15}h^{-1}M_\odot$, 
which means that it is less massive than the most massive high-density cores in SCl~027.
The richest, most elongated, and highest mass component in SCl~019 is 
centred on the group Gr5278 (first component in Table~\ref{tab:comp19}).
This group corresponds to the Zwicky cluster, \object{Zw1215.1+0400}
\citep{1961cgcg.book.....Z}.
The first component also hosts another rich cluster, \object{A1516}. 
The largest extent of this component is $\approx 65$~\Mpc.
Another rich component (2) in SCl~019 hosts two rich groups, one
of them in galaxy cluster \object{A1424}. 

Four components in SCl~019 have median densities $D8 < 7$, they 
are poorer and with total masses lower than $1\times~10^{15}h^{-1}M_\odot$.  
Comparison of Figs. \ref{fig:mclust27} and \ref{fig:mclust19} 
shows that the components in SCl~019 have a higher variety of shapes and
orientations (in the sky distribution) than those in SCl~027.
Components are flattened along the line of sight.

\begin{table*}%[ht]
\setlength{\tabcolsep}{4pt}
\caption{Data of the SCl~019 components.}
\center
\begin{tabular}{rrrrrrrrrrrrr} 
\hline\hline  
(1)&(2)&(3)&(4)& (5)&(6)&(7)&(8)&(9)&(10)&(11)&(12)&(13)\\      
\hline 
 $Nr$ & $N_{\mathrm{gr}}$ & $N_{\mathrm{10}}$ & $N_{\mathrm{1}}$ 
 & $M_{\mathrm{dyn}}$ &  $M^{\mathrm{g}}_{\mathrm{tot}}$ & $dx$ & $dy$ & $dz$ & $dx/dy$ & $dz/dy$ & $D8$& ID\\
 & && & [$10^{15}h^{-1}M_\odot$]& [$10^{15}h^{-1}M_\odot$] &$h^{-1}$ Mpc&$h^{-1}$ Mpc&$h^{-1}$ Mpc & & & \\ 
\hline
1 &288& 15& 174& 2.9&3.2&28.6& 64.8&25.3 &0.4& 0.4& 7.0& \object{Zw1215.1+0400} \\
2 &101&  4&  57& 1.1&1.2&11.6& 16.1&17.8 &0.7& 1.1& 7.6& \object{A1424}\\
3 & 81&  3&  45&  .7& .8&10.0& 20.0&30.3 &0.4& 1.5& 5.7& \\
4 & 63&  4&  35&  .5& .6&14.0& 25.0&11.9 &0.5& 0.4& 5.5& \\
5 & 44&  2&  27&  .5& .5&10.2& 12.4&12.6 &0.8& 1.0& 6.6& \\
6 & 83&  1&  43&  .3& .3& 9.1& 20.7&10.7 &0.4& 0.5& 5.9& \\
\hline                
\label{tab:comp19}  
\end{tabular}\\
\tablefoot{
Columns are the same as in Table~\ref{tab:comp27}.
}
\end{table*}

\begin{table}%[ht]
\setlength{\tabcolsep}{4pt}
\caption{Masses and radii of the central regions of the SCl~019 components.}
\center
\begin{tabular}{rrrrrr} 
\hline\hline  
(1)&(2)&(3)&(4)&(5)&(6)\\      
\hline 
 $Epoch$ & $N_{\mathrm{gr}}$ & $N_{\mathrm{1}}$ & $M_{\mathrm{dyn}}$ &  $M^{\mathrm{g}}_{\mathrm{tot}}$ & $R$ \\
  && & [$10^{15}h^{-1}M_\odot$]& [$10^{15}h^{-1}M_\odot$] &$h^{-1}$ Mpc\\ 
\hline
\multicolumn{2}{l}{Gr5278 region}&&&&\\
 $T$   & 32 & 57   & 1.61& 1.77  & 7.5 \\
 $FC$  & 48 & 95   & 1.84& 2.02  & 9.0 \\
\hline          
\multicolumn{2}{l}{A1424 region}&&&&\\
 $T$   & 20 & 34   & 0.74& 0.82  & 6.0 \\
 $FC$  & 27 & 45   & 0.82& 0.94  & 7.0 \\
\hline
\label{tab:D8prop19}  
\end{tabular}\\
\tablefoot{
Columns are as follows:
(1): Epoch (T: turnaround, FC: future collapse);
(2): the number of groups in a region;
(3): the number of single galaxies in a region;
(4): the dynamical mass of a region;
(5): the total mass of a region;
(6): radius of the region.
}
\end{table}

\begin{figure}%[ht]
\centering
\resizebox{0.445\textwidth}{!}{\includegraphics[angle=0]{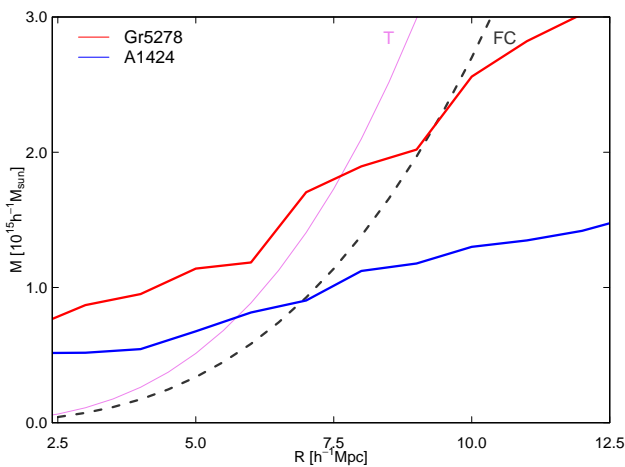}}
\caption{
Mass-radius relation for the high-density cores 
in the supercluster SCl~019. 
Violet line show the embedded mass $M$ versus radius of a sphere $R$ 
for the turnaround $T$, and the dashed grey line shows the future collapse  $FC$.
The red line shows the embedded mass versus radius for the region around the cluster Gr5278, 
the blue line shows the region around the cluster A1424. 
}
\label{fig:mr19}
\end{figure}

Next we analyse the group content and mass distribution
around the rich galaxy clusters Gr5278 and A1424
in the high-density cores of SCl~019 and compare 
it with the predictions of the spherical collapse model
(Fig.~\ref{fig:mr19} and Table~\ref{tab:D8prop19}). 

{\it Gr5278 region.}
Two rich clusters,  Gr5278 
with a mass of  $0.6\times~10^{15}h^{-1}M_\odot$
and \object{A1516} with a mass of $0.4\times~10^{15}h^{-1}M_\odot$ lie in this region.
They are both X-ray clusters
\citep{2007A&A...461..397P, 2011A&A...534A.109P}. The cluster Gr5278 is the richest and
most massive cluster in SCl~019.
Comparison with the spherical collapse model 
shows that the size of the possible (now or in the future) collapsing 
region is 
smaller than $10$~\Mpc, as we found also for the supercluster SCl~027.

The mass-to light ratio of the central region of the component is higher than
for the full supercluster, $\approx 350$~$h\,M_\odot$/$L_\odot$, because of the very massive
cluster in the region. 
The fraction of single galaxies in the turnaround region is 8\%,
and in the future collapse region 12\%.

{\it A1424 region.}
This component is separated from the Gr5278 component by a small underdense region.
The cluster A1424 is located at the outskirts of the component,
but since this cluster is the most massive cluster in the component, we
analysed the mass distribution around it.
Table~\ref{tab:D8prop19} and Fig.~\ref{fig:mr19} 
suggest that the region around the 
cluster A1424 with a mass of about $0.8\times~10^{15}h^{-1}M_\odot$
and a radius of about $6$~\Mpc\ may be  collapsing. 
The radius of the future collapse region is about $7$~\Mpc.

In the A1424 region the fraction of single galaxies in both the turnaround 
and the future collapse regions is approximately 12\%.
In SCl~019 $60-75$~\% of the total mass in (now or in the future)
collapsing regions comes from  galaxy groups with
at least ten member galaxies. In the Gr5278 region this fraction is the highest,
75\%. In this supercluster the fraction of the total mass that
comes from faint groups represented by 
single galaxies also decreases towards the region centres.

\subsubsection{Poor superclusters from the SGW}
\label{sect:poor}

Poor superclusters from the SGW are listed in Tables~\ref{tab:scldata}
and \ref{tab:masses}. 
The poor supercluster SCl~499 with its two X-ray clusters
(\object{A1205} and \object{A1238}) corresponds to the X-ray supercluster RXSCJ1116+0206
in the catalogue of \citet{2013MNRAS.429.3272C}.    
The superclusters SCl~499 and SCl~1109 are very elongated, 
and we only studied the mass distribution in SCl~319.
We calculated masses in spheres of increasing
radii centred on the cluster \object{A1066} with a mass of about  
$0.4\times~10^{15}h^{-1}M_\odot$. We found that 
the region that embeds three rich galaxy groups
and has a total mass of $0.69\times~10^{15}h^{-1}M_\odot$ and a radius of $5$~\Mpc\
may have reached the turnaround and started to collapse. 
The region with
a total mass of $0.74\times~10^{15}h^{-1}M_\odot$  
and a radius of $R \approx 6.5$~\Mpc\
(and perhaps the whole supercluster SCl~319)
will collapse in the future.

%%%%%%%%%%%%%%%%%%%%%%%%%%%%%%%%%%%%%%%%%%%%%%%%%%%%%%%%%%%%%%%%%%%%%%%%%

\section{Discussion and summary}
\label{sect:summary}

{\it Masses of the SGW superclusters.} 
We calculated the masses of the SGW superclusters using 
the dynamical masses of galaxy groups as one mass estimate, 
and the stellar masses of the main galaxies in groups 
to obtain the masses of galaxy groups as another 
mass estimate.
The stellar masses of the main galaxies of groups
have previously been  used  to determine the mass
of the BOSS Great Wall superclusters \citep{2016A&A...588L...4L}. 
This approach is especially promising in cases when the
mass estimates of galaxy groups in superclusters are not available,
like in distant superclusters. 

We found  that the two mass estimators agree well. The main uncertainty of
supercluster masses comes from missing groups and clusters,
and from the mass estimate of faint groups.
The bias between the masses of superclusters determined using group masses
and the total mass of the supercluster have been studied from simulations
\citep{2014A&A...567A.144C}. They showed that the bias between masses
depends on the richness of groups; it is lower when low-mass groups 
are used to construct the supercluster catalogues. 
We found that the bias factor (the ratio of the dynamical mass of the supercluster
and the total mass of the supercluster) is 
of about 
$1.4$, which agrees well with results reported by 
 \citet{2014A&A...567A.144C}
based on simulations, considering that they used higher mass groups to determine
superclusters. This is also similar to what \citet{2015A&A...580A..69E} found for the supercluster 
A2142.

The masses of the SGW superclusters and their high-density core regions
are in the  mass range 
of other observed superclusters of average richness
\citep[see e.g.  ][]{2008MNRAS.390..289J,
2011A&A...532A..57S, 2016A&A...592A...6P, 2015A&A...580A..69E}.
They are lower than the masses of very rich  superclusters such as the
Shapley supercluster, the Corona Borealis supercluster,
and other very rich superclusters 
\citep{2000AJ....120..523R, 2006A&A...447..133P, 2006A&A...445..819R, 
2014MNRAS.441.1601P, 2015MNRAS.453..868O}.
The mass-to-light ratios of the SGW superclusters, 
$M/L \approx 300$~$h\,M_\odot$/$L_\odot$, 
are close to the values of the mass-to-light ratios
for other rich superclusters 
\citep{2004A&A...422..407G, 2011A&A...532A..57S, 2015A&A...580A..69E}.
Moreover, the components in rich SGW superclusters have masses that are comparable to the
masses of simulated superclusters of average richness
\citep{2014A&A...567A.144C, 2009MNRAS.399...97A}.
The masses of rich SGW superclusters are of the same order as the high end of
simulated supercluster masses.

The total volume and mass of the SGW were estimated in \citet{2011MNRAS.417.2938S}.
The authors obtained that the volume of the SGW is 
$7.2\times10^{5}(h^{-1}\mathrm{Mpc})^3$, the effective radius $55$~\Mpc,
and the total mass of the SGW is
$1.2\times~10^{17}h^{-1}M_\odot$. These values are
higher than we obtained, 
the difference comes from the different way of estimating the 
volume and mass of the supercluster.
\citet{2011MNRAS.417.2938S} assumed that the shape of the SGW 
can be approximated with a sphere, which led to overestimation of
the volume and mass of the SGW in comparison with our study.
We found that the effective radius of the full SGW 
(radius of a sphere with the volume of the SGW) is approximately $22$~\Mpc,
significantly smaller than estimated by \citet{2011MNRAS.417.2938S}.

%%%%%%%%%%%%%%%
{\it Mass distribution in the high-density cores of superclusters.} 
We analysed the radial mass distribution in the high-density cores of
rich SGW superclusters centred on
rich galaxy clusters and compared it with the predictions 
of the spherical collapse model. This comparison showed 
that the central
regions of the components with radii up to approximately
$6-7$~\Mpc\ may be collapsing now or in the future.
This limit
corresponds to the size of the shortest axis of very elongated
regions, the size of possibly collapsing regions along the
longest axes of systems may be
larger.
The analysis of the correlations between velocities of galaxy clusters
from simulations showed that up to separations of $10$~\Mpc\
clusters approach each other and their attraction dominates
the bulk motions \citep{1994ApJ...437L..51C}. 
This agrees well with
the minimum size of possibly collapsing regions in supercluster components.

The collapsing high-density cores have been studied 
in  the Shapley and the Corona Borealis superclusters
\citep{2000AJ....120..523R, 2014MNRAS.441.1601P,2015A&A...575L..14C}, 
in the Perseus-Pisces supercluster \citep{2001A&A...378..345H, 2015A&A...577A.144T},
in the A2199 supercluster, the member
of the Hercules supercluster \citep{2001AJ....122.2222E, 2002AJ....124.1266R},
in the SC0028-0005 supercluster \citep{2015MNRAS.453..868O},
and in the A2142 supercluster \citep{2015A&A...580A..69E, 2015A&A...581A.135G}.
In these studies several methods were used to estimate the size and 
mass of the collapsing regions, based on overdensity criteria
and dynamical criteria
\citep[see e.g.   ][]{2000AJ....120..523R, 
2007MNRAS.376.1577D, 2015A&A...575L..14C}.
Comparison of the methods shows that radii of the
collapsing regions obtained with different methods agree well \citep{2000AJ....120..523R,
2014MNRAS.441.1601P, 2015A&A...575L..14C}.
These studies have shown that the sizes of collapsing cores
of superclusters typically do not exceed approximately $10$~\Mpc, which is also
what we found for the central regions of the SGW supercluster components.
According to \citet{2000AJ....120..523R}
and \citet{2014MNRAS.441.1601P},
very massive collapsing cores with a number of rich galaxy clusters
in the Shapley supercluster and in the
Corona Borealis supercluster are larger and more massive.
\citet{2014MNRAS.441.1601P} noted that the collapsing core
in the Corona Borealis supercluster has this size only if there is
a reasonable amount of intercluster mass.
Full very rich and massive superclusters are not bound systems
\citep{2013MNRAS.429.3272C, 2015A&A...575L..14C}, 
as also suggested in our study of the SGW superclusters.
Individual components in the SGW superclusters may form
separate superclusters in the future.

\begin{figure}%[ht]
\centering
\resizebox{0.445\textwidth}{!}{\includegraphics[angle=0]{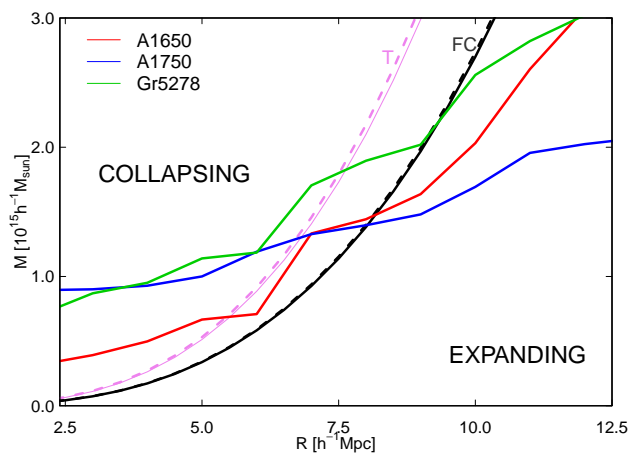}}
\caption{
Mass-radius relation for two $\Omega_{\mathrm{m}}$ values.
The violet lines denote the turnaround $T$ and the black lines denote
the future collapse $FC$.
The dashed lines correspond to $\Omega_{\mathrm{m}} = 0.3$ and the solid lines 
to $\Omega_{\mathrm{m}} = 0.27$.
Other lines denote high-density cores in SCl~027 and SCl~019, as shown in the plot.
}
\label{fig:mromega}
\end{figure}

The mutual distances between the component centres 
in the supercluster SCl~027 are approximately $35$~\Mpc\ 
and $50$~\Mpc, and in the supercluster SCl~019 approximately $20$~\Mpc. 
Numerical simulations show that the positions of the high-density
peaks in the galaxy distribution in the early Universe are fixed by 
the processes  during or just after the inflation and do not change much 
during the cosmic evolution, only the amplitude of
over- and underdensities grows with time 
\citep[][and references therein]{1988Natur.334..129K, 
1996Natur.380..603B, 2009LNP...665..291V, 2011A&A...531A.149S}.
Observational data of the  movements of galaxies 
in the nearby Laniakea and Arrowhead superclusters show that 
galaxy velocities are directed towards supercluster axis
\citep{2014Natur.513...71T, 2015ApJ...812...17P}, and galaxies are
moving from underdense regions towards high-density regions.
This process may accelerate the collapse of individual cores,
but it is unlikely that the cores will approach each other
and merge.

Several studies have shown that the values of the overdensity for turnaround
and future collapse only weakly  depend on the exact value of 
$\Omega_{\mathrm{\Lambda}}$
and change only slightly with matter density $\Omega_{\mathrm{m}}$ 
\citep{1991ApJ...374...29L, 2015A&A...575L..14C, 2015A&A...581A.135G}.
For example, for $\Omega_{\mathrm{m}} = 0.27$, as adopted in the present
paper, the overdensity for the turnaround 
$\Delta\rho_{\mathrm{T}} =  13.1$, and for $\Omega_{\mathrm{m}} = 0.3$
\citep{2015arXiv150201589P}, 
$\Delta\rho_{\mathrm{T}} =  12.2$ \citep{2015A&A...581A.135G}. 
In Fig.~\ref{fig:mromega} we show the mass-radius relation for 
$\Omega_{\mathrm{m}} = 0.27$ and $\Omega_{\mathrm{m}} = 0.3$
together with the radial mass distribution in the SGW superclusters.
The differences between overdensity values for various 
$\Omega_{\mathrm{m}}$ values are much
smaller than the uncertainties of the mass estimates of supercluster high-density
core regions \citep[for overdensity values see the tables in][]
{2015A&A...575L..14C, 2015A&A...581A.135G}. 
Therefore it is questionable to apply the turnaround as a cosmological
test in supercluster scales, as proposed, for example, by \citet{2014JCAP...09..020P}.
More precise mass estimates for a larger number of supercluster (cores) 
are needed for this.

{\it The structure of rich SGW superclusters and supercluster
morphology.} 
Normal mixture modelling showed that the richest SGW superclusters
consist of a number of components. The richest components
hosts rich galaxy clusters; they are  the high-density cores of superclusters.
Components are very elongated in the sky distribution, 
their short axes point along the line of sight. This may be due to their 
real shape, but the flatness of supercluster components 
may also be enhanced by the large-scale velocity field
\citep{1987MNRAS.227....1K, 1994ApJ...425..382G}.

The distribution of components in superclusters reflects the overall 
morphology of superclusters. 
In SCl~027 the components lie along the main body of the supercluster
described as being of filament morphology \citep{2011A&A...532A...5E}.
SCl~019 was described morphologically as having a rich and 
complex inner structure with many galaxy chains connecting
galaxy clusters in the supercluster 
 \citep[spider-type morphology;][]{2011A&A...532A...5E}. 
This is reflected in the much less regular distribution of the
components  in SCl~019
in comparison with SCl~027.
Earlier studies have shown that the morphology of superclusters
and the properties of galaxies and groups  in them are related. Superclusters
of filament morphology host a higher fraction of red passive galaxies
than the superclusters of spider morphology, where  the galaxy groups
also have a higher amount of substructure \citep{2012A&A...542A..36E,
2014A&A...562A..87E}. These 
differences may be related with the dynamical state of the superclusters and galaxy
groups in them,
spider-type superclusters being dynamically younger and more active
than filament-type superclusters. 
The detailed study of the properties of galaxies and galaxy groups
in the richest SGW superclusters and their high-density cores may 
give us an insight into their dynamical state.

We summarise the main results as follows:

We determined the masses of the SGW superclusters using dynamical masses
of groups and stellar masses of the main galaxies in groups, and
found a good agreement between mass estimates.
The lower limit of the total mass of the SGW is 
approximately $M = 2.5 \times~10^{16}h^{-1}M_\odot$.

We applied normal mixture modelling to study the structure of
superclusters and identify their high-density cores. 
This is a new and promising
approach to study the structure of
superclusters and detect their high-density cores.
 
The richest SGW superclusters consists of several very elongated
high-density cores with masses of the richest
components of up to  $6 \times~10^{15}h^{-1}M_\odot$ 
and sizes of up to $65$~\Mpc. Their short axes lie
approximately along the line of sight, and their sizes
are typically smaller then $20$~\Mpc.
 
The core regions of the components with radii smaller than $8$~\Mpc\ 
may already be collapsing. This is probably
only the shortest size of the collapsing regions.

The study of the properties of galaxies and galaxy groups in 
different regions in superclusters may provide 
information about environmental effects that shape the
galaxy properties in superclusters, and about the regions themselves.
We plan to continue the study of the dynamical state of galaxy superclusters
from observations and simulations, as well as 
the  properties of galaxies and their systems (groups, clusters, and filaments) 
in dynamically evolving environment of superclusters.

\section*{Acknowledgments}

We thank the referee for valuable comments and suggestions.

We are pleased to thank the SDSS Team for the publicly available data
releases.  Funding for the Sloan Digital Sky Survey (SDSS) and SDSS-II has been
provided by the Alfred P. Sloan Foundation, the Participating Institutions,
the National Science Foundation, the U.S.  Department of Energy, the
National Aeronautics and Space Administration, the Japanese Monbukagakusho,
and the Max Planck Society, and the Higher Education Funding Council for
England.  The SDSS Web site is \texttt{http://www.sdss.org/}.
The SDSS is managed by the Astrophysical Research Consortium (ARC) for the
Participating Institutions.  The Participating Institutions are the American
Museum of Natural History, Astrophysical Institute Potsdam, University of
Basel, University of Cambridge, Case Western Reserve University, The
University of Chicago, Drexel University, Fermilab, the Institute for
Advanced Study, the Japan Participation Group,  Johns Hopkins University,
the Joint Institute for Nuclear Astrophysics, the Kavli Institute for
Particle Astrophysics and Cosmology, the Korean Scientist Group, the Chinese
Academy of Sciences (LAMOST), Los Alamos National Laboratory, the
Max-Planck-Institute for Astronomy (MPIA), the Max-Planck-Institute for
Astrophysics (MPA), New Mexico State University, Ohio State University,
University of Pittsburgh, University of Portsmouth, Princeton University,
the United States Naval Observatory, and the University of Washington.

In this work we used R statistical environment
\citep{ig96}.

We acknowledge the support by the Estonian Research Council grants IUT26-2, IUT40-2,
and by the
Centre of Excellence “Dark side of the Universe” (TK133) financed by the
European Union through the European Regional Development Fund, and by
 ICRAnet through a professorship for Jaan Einasto.
HL is supported by Turku University Foundation. 

%\end{acknowledgements}

\bibliographystyle{aa}
\bibliography{sgw.bib}

\end{document}